\pdfoutput=1
\documentclass[aps,prl,10pt,twocolumn,superscriptaddress]{revtex4}

\usepackage{mathrsfs, amssymb, amsmath, amsfonts, txfonts, latexsym, graphicx}  
\usepackage[utf8]{inputenc}
	\usepackage[
      colorlinks=true,
      linkcolor=blue,
      urlcolor=cyan,
      filecolor=magenta,
      citecolor=red,
      pdfstartview=FitV,
      hyperfootnotes=false,
      pdftitle={Carroll geodesics},
        pdfauthor={Luca Ciambelli, Daniel Grumiller},
        pdfsubject={Carroll geodesics},
        pdfkeywords={Carroll geodesics},
        bookmarksopen=true
      ]{hyperref}
\usepackage[all]{hypcap}
\usepackage{soul}

\newcommand{\eq}[2]{\begin{equation} #1 \label{#2} \end{equation}}
\DeclareMathOperator{\extdm}{d}
\newcommand{\extd}{\extdm \!}

\DeclareFontFamily{OT1}{rsfs}{}
\DeclareFontShape{OT1}{rsfs}{m}{n}{ <-7> rsfs5 <7-10> rsfs7 <10->rsfs10}{} 
\DeclareMathAlphabet{\mycal}{OT1}{rsfs}{m}{n}

\newcommand{\com}{\frac{1}{b^2}} 

\begin{document}


\title{Carroll geodesics}

\author{Luca Ciambelli}
\affiliation{Perimeter Institute for Theoretical Physics, 31 Caroline St. N., Waterloo ON, Canada, N2L 2Y5}
\email{ciambelli.luca@gmail.com}

\author{Daniel Grumiller}
\affiliation{Perimeter Institute for Theoretical Physics, 31 Caroline St. N., Waterloo ON, Canada, N2L 2Y5}
\affiliation{Institute for Theoretical Physics, TU Wien, Wiedner Hauptstrasse 8–10/136, A-1040 Vienna,
Austria}
\email{grumil@hep.itp.tuwien.ac.at}

\date{\today}

\begin{abstract}
Using effective field theory methods, we derive the Carrollian analog of the geodesic action. We find that it contains both ``electric'' and ``magnetic'' contributions that are in general coupled to each other. The equations of motion descending from this action are the Carrollian pendant of geodesics, allowing surprisingly rich dynamics. As an example, we derive Carrollian geodesics on a Carroll--Schwarzschild background and discover an effective potential similar to the one appearing in geodesics on Schwarzschild backgrounds. However, the Newton term in the potential turns out to depend on the Carroll particle's energy. As a consequence, there is only one circular orbit localized at the Carroll extremal surface, and this orbit is unstable. For large impact parameters, the deflection angle is half the value of the general relativistic light-bending result. For impact parameters slightly bigger than the Schwarzschild radius, orbits wind around the Carroll extremal surface. For small impact parameters, geodesics get reflected by the Carroll black hole, which acts as a 
perfect mirror. 
\end{abstract}

\maketitle


\section{Introduction}

Geometry and physics have a long-lasting relationship that has evolved considerably over time. In the days of Plato's school, whose entrance door allegedly had the engraving ``Let no one ignorant of geometry enter,'' space was Euclidean, and space and time were absolute. The only symmetries were space- and time-translations and spatial rotations. Nowadays, such spacetimes are referred to as ``Aristotelian''. 

The first substantial change in geometry came from physics: Galilei dropped absolute space and introduced relativity, using gedanken- and real experiments with ships. In modern jargon, the new symmetries introduced by Galilei are called boosts. Space was now relative, though time remained absolute in these Galilean spacetimes. 

The second significant change came from mathematics: Riemann (after pioneering work by Lobachevsky, Bolyai, Gauss, and others) dropped Euclid's fifth axiom and established curved spaces with signature $(+,+,\dots,+)$. Symmetries were then formalized in terms of Killing vectors. 

The third impactful change came from mathematics and physics in the wake of special (and later general) relativity: space and time were merged into a single entity, spacetime, requiring metrics with a pseudo-Riemannian signature $(-,+,\dots,+)$. The Galilean boosts became Lorentzian, and neither space nor time remained absolute.

Currently, we experience the Carrollian revolution started by L\'evy-Leblond more than half a century ago \cite{Levy1965,SenGupta1966OnAA}. This fourth major change in our understanding of geometry permits zeros in the metric signature: Carrollian spacetimes have a degenerate signature of $(0,+,\dots,+)$. The simplest example of such a spacetime is the vanishing speed of light limit of the Minkowski metric $\extd s^2=\lim_{c\to 0}(-c^2\,\extd t^2+\delta_{ij}\,\extd x^i\extd x^j)=\delta_{ij}\,\extd x^i\extd x^j$. Carrollian time is relative, but Carrollian space is absolute. Moreover, to characterize a Carrollian spacetime, in addition to a Carroll metric $h_{\mu\nu}$ of signature $(0,+,\dots,+)$, we need a Carroll vector $v^\mu$ that lies in the kernel of the metric, $v^\mu\,h_{\mu\nu}=0$. In the example just discussed, the vector field is $v=v^\mu\partial_\mu=\partial_t$ and the Carroll metric is $h_{\mu\nu}=\delta_{ij}\,\delta^i_\mu\,\delta^j_\nu$.

A remarkable aspect of Carrollian spacetimes is that they can possess infinite-dimensional isometries, unlike \mbox{(pseudo-)}\-Riemannian spacetimes. Indeed, for the example above, solving the Killing equations $\xi^\mu\partial_\mu h_{\alpha\beta}+h_{\alpha\mu}\partial_\beta\xi^\mu+h_{\mu\beta}\partial_\alpha\xi^\mu=0=v^\mu\partial_\mu\,\xi^\nu-\xi^\mu\partial_\mu\,v^\nu$ yields Euclidean translations and rotations, and additionally an infinite set of Killing vectors $\xi=f(x^i)\,\partial_t$ that preserve the Carroll metric $h_{\mu\nu}$ and the Carroll vector $v^\mu$.

It took the physics community a while to figure out that Carrollian spacetimes are applicable to anything --- contrary to the Galilean limit, where the speed of light tends to infinity, there is no everyday intuition associated with the Carrollian limit. However, in the past decade, several physics applications of Carrollian symmetries have emerged.

Perhaps the clearest one is associated with the structure of null infinity, see e.g.~\cite{Ciambelli:2018wre, Figueroa-OFarrill:2021sxz, Herfray:2021qmp,Mittal:2022ywl, Campoleoni:2023fug}. Formally, the metric at null infinity is $0\cdot\extd u^2+\extd\Omega^2$, where $u$ is retarded (or advanced) time and $\extd\Omega^2$ the metric of the celestial sphere. The infinite-dimensional Carrollian symmetries are the supertranslations discovered by Bondi, van~der~Burgh, Metzner, and Sachs (BMS) \cite{Bondi:1962,Sachs:1962}. So null infinity in asymptotically flat spacetimes naturally carries a Carrollian structure, and the BMS symmetries are equivalent to (conformal) Carrollian symmetries \cite{Duval:2014uoa, Duval:2014uva, Duval:2014lpa}. The relation between BMS and conformal Carrollian symmetries paved the way for the Carrollian approach to flat space holography, which so far has been studied mostly in three spacetime dimensions \cite{Barnich:2006av,Bagchi:2010zz,Bagchi:2012yk,Barnich:2012xq,Bagchi:2012xr,Barnich:2012rz,Bagchi:2013lma,Bagchi:2014iea,Barnich:2015mui,Campoleoni:2015qrh,Bagchi:2015wna,Bagchi:2016bcd,Jiang:2017ecm,Grumiller:2019xna,Apolo:2020bld} and more recently in four spacetime dimensions ~\cite{Ciambelli:2018wre, Figueroa-OFarrill:2021sxz, Herfray:2021qmp, Donnay:2022aba, Bagchi:2022emh,Campoleoni:2022wmf, Donnay:2022wvx, Mittal:2022ywl, Campoleoni:2023fug, Salzer:2023jqv, Bagchi:2023fbj, Saha:2023hsl}. Other physics applications of Carrollian symmetries include the description of null hypersurfaces \cite{Penna:2015gza, Penna:2018gfx, Donnay:2019jiz, Ciambelli:2019lap, Redondo-Yuste:2022czg, Freidel:2022vjq, Gray:2022svz, Ciambelli:2023mir, Ciambelli:2023mvj}, the fluid/gravity correspondence \cite{deBoer:2017ing, Ciambelli:2018xat, Campoleoni:2018ltl, Ciambelli:2020eba, Ciambelli:2020ftk, Freidel:2022bai, Petkou:2022bmz}, Carrollian algebra, scalar fields and particles \cite{Bergshoeff:2014jla, Henneaux:2021yzg, Bagchi:2022eav, Bekaert:2022oeh, Bergshoeff:2022eog, Rivera-Betancour:2022lkc, Ekiz:2022wbi, Baiguera:2022lsw, Kasikci:2023tvs, Casalbuoni:2023bbh, Cerdeira:2023ztm, Kamenshchik:2023kxi, Zhang:2023jbi}, tensionless strings \cite{Bagchi:2015nca, Bagchi:2019cay,Bagchi:2020ats, Fursaev:2023lxq, Fursaev:2023oep}, cosmology \cite{deBoer:2021jej}, Hall effects \cite{Marsot:2022imf}, fractons \cite{Bidussi:2021nmp, Figueroa-OFarrill:2023vbj, Figueroa-OFarrill:2023qty, Perez:2023uwt}, flat bands \cite{Bagchi:2022eui}, Bjorken flow \cite{Bagchi:2023ysc}, supersymmetry and supergravity \cite{Ravera:2019ize, Ali:2019jjp, Ravera:2022buz, Kasikci:2023zdn}. See \cite{deBoer:2023fnj,Ciambelli:2023xqk} for more references.

Given the success of general relativity, which geometrically can be understood as emerging from gauging the Poincar\'e-algebra, and the ubiquitousness of Carroll symmetries, it is natural to gauge the Carroll algebra \cite{Hartong:2015xda} and establish Carroll gravity theories \cite{Henneaux:1979vn, Bergshoeff:2017btm,Ciambelli:2018ojf,Matulich:2019cdo,Grumiller:2020elf,Gomis:2020wxp,Perez:2021abf,Hansen:2021fxi,deBoer:2021jej,Concha:2021jnn,Figueroa-OFarrill:2022mcy,Campoleoni:2022ebj, Miskovic:2023zfz}, which may exhibit Carroll black hole solutions \cite{Ecker:2023uwm}.  

So far, the only discussion in the literature on how to probe Carrollian spacetimes with test particles is in \cite{deBoer:2023fnj}, where the approach is to take the Carrollian limit of the relativistic geodesic equation. There, it was found that in the limit  particles either cannot move or have zero energy.

To paraphrase Wheeler, we know already the way in which ``matter tells Carroll geometry how to curve''. The main purpose of our Letter is to establish how ``Carroll geometry tells matter how to move''. In other words, our goal is to derive the intrinsic Carrollian version of geodesics.

Our main conclusion is that our intrinsic analysis shows that Carroll particles following Carrollian geodesics can move, with arbitrary energy. We shall demonstrate this result first in full generality and then by means of a pertinent example, a Carroll test particle moving on the background of a Carroll--Schwarzschild black hole.  Our conclusions are supported by the analysis of a companion paper \cite{Ciambelli:2023xqk} that analyzes Carroll scalar fields and also finds non-trivial dynamics.
  
\section{Carrollian geometry basics}

We review salient features of Carrollian geometry, see \cite{Hartong:2015xda, Ciambelli:2019lap, Hansen:2021fxi, Ecker:2023uwm} for more details. In addition to the Carroll metric $h_{\mu\nu}$ and the Carroll vector $v^\mu$, we introduce the dual of the Carroll vector, $E_\mu$, known in the literature as ``clock 1-form'' or ``Ehresmann connection''. This allows defining the projector $h_{\mu}{}^\nu=\delta^\nu_\mu-E_\mu v^\nu$. The main properties of these quantities are 
\eq{
h_{\mu\nu} v^\nu = v^\nu h_{\nu}{}^\mu=h_\nu{}^\mu E_\mu=0\quad E_\mu v^\mu=1\quad h_\mu{}^\alpha h_\alpha{}^\nu=h_\mu{}^\nu\,.
}{eq:cg1}

We split the coordinates into perpendicular and parallel components ($\Omega$ is a positive scale factor to be fixed),
\eq{
x^\mu = x^\mu_\perp+\Omega\,v^\mu x_\parallel\qquad x^\mu_\perp:=x^\nu h_\nu{}^\mu \qquad x_\parallel:=\frac{E_\mu x^\mu}{\Omega}\,.
}{eq:cg2}
We demand an analogous split for the coordinate differentials,
\eq{
\extd x^\mu=\delta^\mu_\nu \extd x^\nu=\extd x^\nu(h_\nu{}^\mu+E_\nu v^\mu)\stackrel{!}{=}\extd x^\nu_\perp h_\nu{}^\mu+\Omega v^\mu\extd x_\parallel
}{eq:cg100}
where the first two equalities hold by definition, and the last one imposes non-trivial requirements: $v^\mu$ changes only parallel to itself, $\extd{v}^\mu=v^\mu\extd\alpha$ with some scalar $\alpha$, and we exploit local Carroll boosts $E_\mu\to{E}_\mu-\lambda_\mu$ to impose $\extd{E_\mu}=-E_\mu\,\extd\alpha$, consistent with $\extd\,(E_\mu{v}^\mu)=0$. Together, these conditions imply $\extd\,h_\nu{}^\mu=0$ and thus $\extd{x}^\mu_\perp=\extd{x}^\nu_\perp \, h_\nu{}^\mu$. The remaining term in \eqref{eq:cg100} is compatible, provided we fix the scale factor $\Omega=\exp{(-\alpha)}$.

These requirements are invariant under the Carroll diffeomorphisms introduced in \cite{Ciambelli:2018wre}, see \cite{Ciambelli:2019lap}. From these data, we extract two Carroll-diffeomorphism invariant quantities,
$h_{\mu\nu}\,\extd{x}^\mu_\perp\extd{x}^\nu_\perp$ and $\extd{x}_\parallel$. Both quantities are also invariant under local Carroll boosts.

\section{Carroll geodesics}

The philosophy of effective field theories is to write down the most general action compatible with the field content, the symmetries, and possibly further consistency requirements and then perform a derivative expansion, keeping only the terms with a certain number of derivatives. We apply this scheme to derive an action for Carrollian test particles moving in the background of an arbitrary Carrollian spacetime.

Let us start with symmetries. We require the Carrollian geodesic action to be invariant under worldline reparametrizations and Carroll diffeomorphisms. The field content is $x_\parallel$, $x_\perp^\mu$, and the worldline einbein $e$. Because of translation invariance, the action only should depend explicitly on $\dot x_\parallel$ and $\dot x^\mu_\perp$, where dots denote derivatives with respect to the worldline parameter $\tau$. Keeping only terms with up to two derivatives (and dropping total derivative terms) yields the action~\footnote{In lower dimensions, additional parity-odd terms can appear. In two dimensions the term $e^{-1}\epsilon_{\mu\nu}\dot{x}^\mu_{\perp}v^\nu$ can be added to \eqref{eq:cg3}, and in three dimensions the term $e^{-2}\epsilon_{\mu\nu\sigma}\dot{x}_\perp^\mu\dot{x}_\perp^{\nu}v^\sigma$. The unique (parity even) term with only one derivative, $e^{-1}\dot{x}_\parallel$, is a total derivative term and hence was dropped.}
\eq{
\boxed{
I[x_\parallel,\,x^\mu_\perp,\,e]=\int\extd\tau\,e\,\Big(g_0+g_1e^{-2}h_{\mu\nu}\dot x^\mu_\perp\dot x^\nu_\perp+g_2e^{-2}\dot x_\parallel^2\Big)
}
}{eq:cg3}
where the remaining freedom are the coupling constants $g_i$. The action \eqref{eq:cg3} is the Carrollian analog of the geodesic action and our first key result.

Some remarks are in order. In Carrollian jargon (see, e.g., \cite{deBoer:2023fnj}), the action \eqref{eq:cg3} contains both an ``electric term'' (proportional to $g_2$) and a ``magnetic term'' (proportional to $g_1$). For instance, the term proportional to $g_1$ emerges from the relativistic geodesic Lagrangian $L=e^{-1}g_2g_{\mu\nu}\dot{x}^\mu\dot{x}^\nu$ in the Carroll limit $x^0=ct$, $c\to0$. Similarly, the term proportional to $g_2$ emerges from the relativistic geodesic Hamiltonian $H=\frac{e}{2}p^2$ in the $c\to 0$ limit after rescaling $e\to ec^2$ and $p_0\to E/c$. It is a key aspect of our Carrollian geodesic action \eqref{eq:cg3} that we both have electric and magnetic terms. For typical Carroll spacetimes, such as in our example below, the quantity $x_\parallel$ is (minus) the coordinate time $t$; it transforms as a scalar under Lie variations, $\mathcal{L}_\xi{x}_\parallel=\xi^\mu\partial_\mu{x}_\parallel$.

Variation of the Carroll geodesic action \eqref{eq:cg3} with respect to the einbein yields the constraint
\eq{
e = \pm\sqrt{\frac{g_1}{g_0}\,h_{\mu\nu}\dot x^\mu_\perp\dot x^\nu_\perp+\frac{g_2}{g_0}\dot x^2_\parallel}\,.
}{eq:cg4}
Like for standard geodesics, we assume from now on an affine parametrization where $e=\rm const$. Variation with respect to $x_\parallel$ yields the equation of motion
\eq{
\boxed{
\phantom{\Big(}
\ddot x_\parallel + \Gamma_{\mu\nu}^\parallel\,\dot x_\perp^\mu\dot x_\perp^\nu=0
\phantom{\Big)}
}
}{eq:cg5}
with the Christoffel-like quantity
$\Gamma_{\mu\nu}^\parallel = -\frac{g_1}{2g_2}\,\partial^\parallel h_{\mu\nu}$, 
where we assumed $g_2\neq0$ and defined $\partial^\parallel:=\partial/\partial x_\parallel$. Similarly, variation with respect to $x^\mu_\perp$ yields the equation of motion
\eq{
\boxed{
\phantom{\Big(}
h_{\sigma\mu}\,\ddot x_\perp^\mu + \Gamma_{\mu\nu\sigma}^\perp\,\dot x_\perp^\mu\dot x_\perp^\nu=0
\phantom{\Big)}
}
}{eq:cg8}
with the Christoffel-like quantity
$\Gamma_{\mu\nu\sigma}^\perp = \frac12\,\big(\partial_\mu^\perp h_{\nu\sigma}+\partial_\nu^\perp h_{\mu\sigma}-\partial_\sigma^\perp h_{\mu\nu}\big)$, 
where we defined $\partial^\perp_\mu:=\partial/\partial x^\mu_\perp$. The equations \eqref{eq:cg5}-\eqref{eq:cg8} are the Carrollian geodesic equations, our second main result.

\section{Carroll--Schwarzschild black hole}

In this section, we study in detail Carroll geodesics on a Carroll--Schwarzschild background.

\subsection{Background structure and Carroll geodesic action}

Rather than directly evaluating the Carrollian geodesic equations \eqref{eq:cg5}-\eqref{eq:cg8}, we insert the Carroll--Schwarzschild black hole ($x^\mu=(t,r,\theta,\varphi)$) \cite{Hansen:2021fxi,Perez:2021abf,Ecker:2023uwm}
\begin{align}
h_{\mu\nu}\,\extd x^\mu\extd x^\nu &= \frac{\extd r^2}{1-\frac{2M}{r}}+r^2\,\big(\extd\theta^2+\sin^2\!\theta\,\extd\varphi^2\big)
\label{eq:cg9}\\ 
v^\mu\,\partial_\mu &=  -\frac{1}{\sqrt{1-\frac{2M}{r}}}\,\partial_t\qquad \alpha=-\tfrac12\,\ln\big(1-\tfrac{2M}{r}\big)
\label{eq:cg10}\\
E_\mu\,\extd x^\mu &= -\sqrt{1-\frac{2M}{r}}\,\extd t\qquad\Omega=\sqrt{1-\frac{2M}{r}}
\label{eq:cg11}
\end{align}
as background into the Carroll geodesic action \eqref{eq:cg3}, obtaining
\eq{
I=\int\extd\tau\,e\,\Bigg[g_0+g_1e^{-2}\,\Bigg(\frac{\dot r^2}{1-\frac{2M}{r}}+r^2\dot\varphi^2\Bigg)+g_2e^{-2}\,\dot x_\parallel^2\Bigg]
}{eq:cg12}
where we assumed, without loss of generality, motion in the equatorial plane $\theta=\pi/2$ and $\dot\theta=0$. Since on this background one has $h_\mu{}^\nu=\text{diag}(0,1,1,1)$, we used the identities $x^t_\perp=0$, $x^r_\perp=r$, $x^\theta_\perp=\theta$, $x^\varphi_\perp=\varphi$, and $x_\parallel=-t$. 

\subsection{Geodesic equations and effective potential}

Varying the action \eqref{eq:cg12} with respect to $x_\parallel$ yields
\eq{
\ddot x_\parallel=0\qquad\Rightarrow\qquad x_\parallel = F\tau+x^0_\parallel
}{eq:cg13}
with some integration constants $F$ and $x^0_\parallel$. Variation with respect to $\varphi$ produces another constant of motion, the angular momentum $\ell$,
\eq{
\partial_\tau\big(r^2\dot\varphi\big)=0\qquad\Rightarrow\qquad\dot\varphi=\frac{\ell}{r^2}
}{eq:cg14}

Like for standard geodesics on a Schwarzschild background, it is efficient to avoid varying the geodesic action with respect to $r$ and instead exploit the affine parametrization using \eqref{eq:cg4}
\eq{
\frac{g_1}{g_0}\,\Bigg(\frac{\dot r^2}{1-\frac{2M}{r}}+\frac{\ell^2}{r^2}\Bigg)+\frac{g_2}{g_0}\,F^2=e^2=\rm const.
}{eq:cg15}
Defining the test particle energy (per mass unit)
\eq{
E:=\frac{g_0}{2g_1}\,e^2-\frac{g_2}{2g_1}\,F^2
}{eq:cg16}
allows rewriting \eqref{eq:cg15} as an energy conservation equation
\eq{
\frac{\dot r^2}{2}+V^{\textrm{\tiny eff}}(r) = E
}{eq:cg17}
with the effective potential
\eq{
\boxed{
\phantom{\Big(}
V^{\textrm{\tiny eff}}(r) = \frac{2ME}{r} + \frac{\ell^2}{2r^2} - \frac{M\ell^2}{r^3}\,.
\phantom{\Big)}
}
}{eq:cg18}

Remarkably, there is only one small but significant difference to the effective potential for standard geodesics in a Schwarzschild background: the first (Newtonian) term in \eqref{eq:cg18} depends linearly on the test particle's energy and can have either sign. As we shall see, this difference has drastic consequences for the orbits of test particles. The technical key aspect is that the energy conservation equation \eqref{eq:cg17} factorizes,
\eq{
\dot r^2 = \bigg(1-\frac{2M}{r}\bigg)\,\bigg(\com-\frac{1}{r^2}\bigg)\,\ell^2
}{eq:lalapetz}
and the first factor on the right-hand side vanishes at the locus $r=2M$ corresponding to a Carroll extremal surface (CES) \cite{Ecker:2023uwm}. Here, we introduced the impact parameter
\eq{\boxed{\phantom{\Big(}
b=\frac{\ell}{\sqrt{2E}} 
\phantom{\Big)}}}{eq:angelinajolie}
that determines the asymptotic radial velocity $|\dot r|_{r\to\infty}=\ell/b$ and plays a prominent role in the forthcoming orbit analysis~\footnote{%
The constant of motion $\ell$ is irrelevant as long as $\ell\neq0$. This can be verified by rescaling the affine parameter $\tau$ by a positive constant, which effectively rescales $\ell$ by the inverse of this constant. Thus, only $b$ is relevant. Whenever $b$ is real we assume it is non-negative, without loss of generality.}. 

\subsection{Circular orbits}

We consider next circular orbits, defined by the circularity conditions
\eq{
V^{\textrm{\tiny eff}}=E\qquad\qquad\frac{\extd V^{\textrm{\tiny eff}}}{\extd r}=0\,.
}{eq:cg19}
Evaluating \eqref{eq:cg19} yields two algebraic equations. 
\eq{
  \bigg(1-\frac{2M}{r}\bigg)\,\bigg(\com-\frac{1}{r^2}\bigg)\,\ell^2=0=
  \bigg(\frac{3M}{r^2}-\frac{M}{b^2}-\frac{1}{r}\bigg)\, \frac{\ell^2}{r^2}
}{eq:cg42}
If the test particle is at the CES, $r=2M$, then the first identity above holds automatically, and the second one yields the constraint $b_{\mathrm{circ}}=2M$. Thus, at the CES circular orbits are possible for any non-negative energy. If we assume instead $r\neq 2M$ we get a contradiction (except for the trivial case $E=\ell=0$, which we disregard~\footnote{%
On trivial orbits, $E=\ell=0$, the azimuthal angle and the radius are fixed, $\dot\varphi=0=\dot r$. Thus, the only ``dynamics'' of such orbits consists of time flow, given by \eqref{eq:cg13}. While formally such orbits obey the circularity conditions \eqref{eq:cg19} we refrain from referring to them as ``circular'' due to their triviality.}): The first circularity condition requires $b=r$, the second circularity condition demands $(2M-r)\ell^2/r^4=0$, and together they imply $r=2M$, negating our original assumption.

We have just proven that there are no circular orbits except at the CES. They are unstable with respect to radial perturbations because the second derivative of the effective potential is negative, $\frac{\extd^2V^{\textrm{\tiny eff}}}{\extd r^2}\big|_{r=b=2M} = -\frac{\ell^2}{8M^4} < 0$.

Thus, the linear dependence on energy in the Newton term in the effective potential \eqref{eq:cg18} completely changes the spectrum of circular orbits compared to standard Schwarzschild geodesics. There are no stable circular orbits (and hence no innermost stable circular orbits), and the only unstable circular orbit is located at the CES, $r=2M$. Outside of the CES, no circular motion is possible. Hence, there is no planetary motion, no Kepler's laws, and no perihelion that can be shifted.

\subsection{Deflection angle}

We now derive the deflection angle for a Carroll test particle incoming from infinity and scattered off the Carroll black hole. Combining the equations of motion \eqref{eq:cg14}-\eqref{eq:cg18} yields
\eq{
\frac{\extd\varphi}{\extd r} = \pm \frac{b}{\sqrt{r(r-2M)(r^2-b^2 )}}
}{eq:cg22}
Introducing the integration variable $u=1/r$, we can represent the deflection angle as
\eq{
\varphi_\infty = 2\int\limits_0^{u_0}\frac{b\extd u}{\sqrt{(1-2Mu)(1-b^2u^2)}}
}{eq:cg23}
where $u_0$ is the turning point of the trajectory, defined by the smallest positive root of the condition $\extd u/\extd\varphi=0$. For impact parameter $b>2M$, we have $u_0=1/b$.

For simplicity, consider first the massless case $M=0$. Then the integral \eqref{eq:cg23} is elementary and yields
\eq{\varphi_{\infty}=2\arcsin(b\,u_0) \stackrel{u_0=1/b}{\rightarrow} \pi\,.
}{eq:bradpitt}
As expected on physical grounds, in the massless case, the deflection angle is $\pi$, i.e., the outgoing trajectory is antipodal to the ingoing trajectory.

To obtain the deflection angle to first order in the mass $M$, we differentiate the integrand in \eqref{eq:cg23} with respect to $M$ and set $M$ to zero before integrating. The deflection angle relative to the case without a Carroll black hole,
\eq{
\varphi_\infty-\pi\approx M\,\frac{\partial\varphi_\infty}{\partial M}\Bigg|_{M=0} =\frac{2M}{b}\bigg(1-\sqrt{1-b^2u_0^2}\bigg)\stackrel{u_0=1/b}{\rightarrow} \frac{2M}{b}
}{eq:cg25}
is half the deflection angle of lightlike geodesics on a Schwarzschild background, see e.g.~\cite{waldgeneral}. 

To all orders in the mass, assuming $u_0=1/b$, the result for the deflection angle is given in terms of elliptic integrals,~\footnote{The functions $K$ and $F$ correspond to the Mathematica functions \texttt{EllipticK} and \texttt{EllpiticF}, respectively.}
\eq{
\varphi_\infty = \frac{4}{\sqrt{1+x}}\,\Bigg(K\bigg(\frac{2x}{1+x}\bigg)-F\bigg(\frac\pi4,\,\frac{2x}{1+x}\bigg)\Bigg)
}{eq:cg26}
where $x:=2M/b\in(0,1)$. The small-$x$ expansion ($b\gg 2M$) is $\varphi_\infty=\pi+x+3\pi x^2/16+\dots$ and recovers \eqref{eq:cg25}, while the expansion when $x$ approaches $1$ from below yields $\varphi_\infty=-\sqrt{2}\,\ln(1-x)+\Delta+\dots$, with $\Delta=\sqrt{2}\ln\frac{32}{(1+\sqrt{2})^2}=2.408\dots$ 

\begin{figure}
\begin{center}
\includegraphics[width=0.8\linewidth]{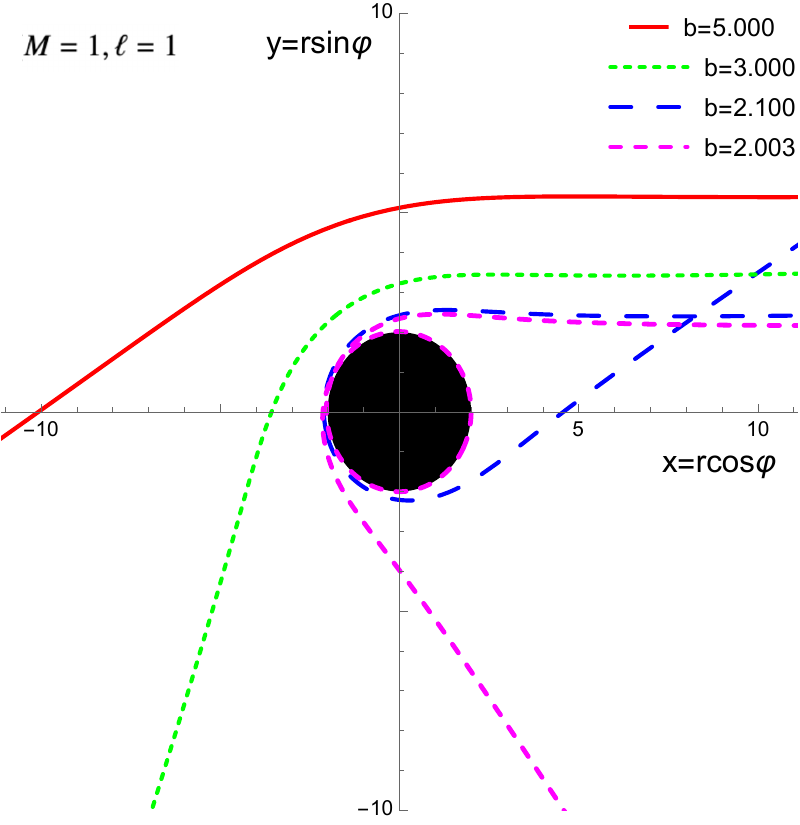}
\end{center}
\caption{For $b>2M$,  Carroll geodesics bend and turn at $u_0=1/b$.}
\label{fig:1}
\end{figure}

The last result means that the orbit winds around the azimuthal direction repeatedly as the limiting case $b\to 2M$ is approached (from above). The approximate value for the impact parameter associated with $n$ windings of the Carroll geodesics around the CES for large $n$ is given by
\eq{
b(n) \approx \frac{2M}{1-e^{-(2\pi n-\Delta)/\sqrt{2}}}\,.
}{eq:cg41}
We have collected some Carroll geodesic orbits in Fig.~\ref{fig:1}.

Note that reality of the velocity \eqref{eq:lalapetz} requires $r^2\geq{b^2}\geq0$ outside the CES. Thus, Carroll particles cannot be placed in the forbidden zone $2M<r<b$.

\subsection{Perfect mirror}

Finally, we consider small impact parameters, $0\leq{b}\leq2M$. At the upper end of the interval, $b=2M$, we have two different types of orbits: the (unstable) circular orbit at $r=2M$ studied above and orbits that arrive tangential to the CES but need infinite worldline time to reach it, as can be seen inserting $b=2M$ into Eq.~\eqref{eq:lalapetz},
\eq{
\frac{2M \,\text{d}r}{(1-2M/r)\,\sqrt{1+2M/r}} = \pm\ell\, \text{d}\tau\,.
}{eq:lalapetzNew}
The case $b=\ell=0$ with $r(0)>2M$ yields a radial geodesic that solves $\ddot r=2ME/r^2$ and bounces back from the CES.

\begin{figure}
\begin{center}
\includegraphics[width=0.8\linewidth]{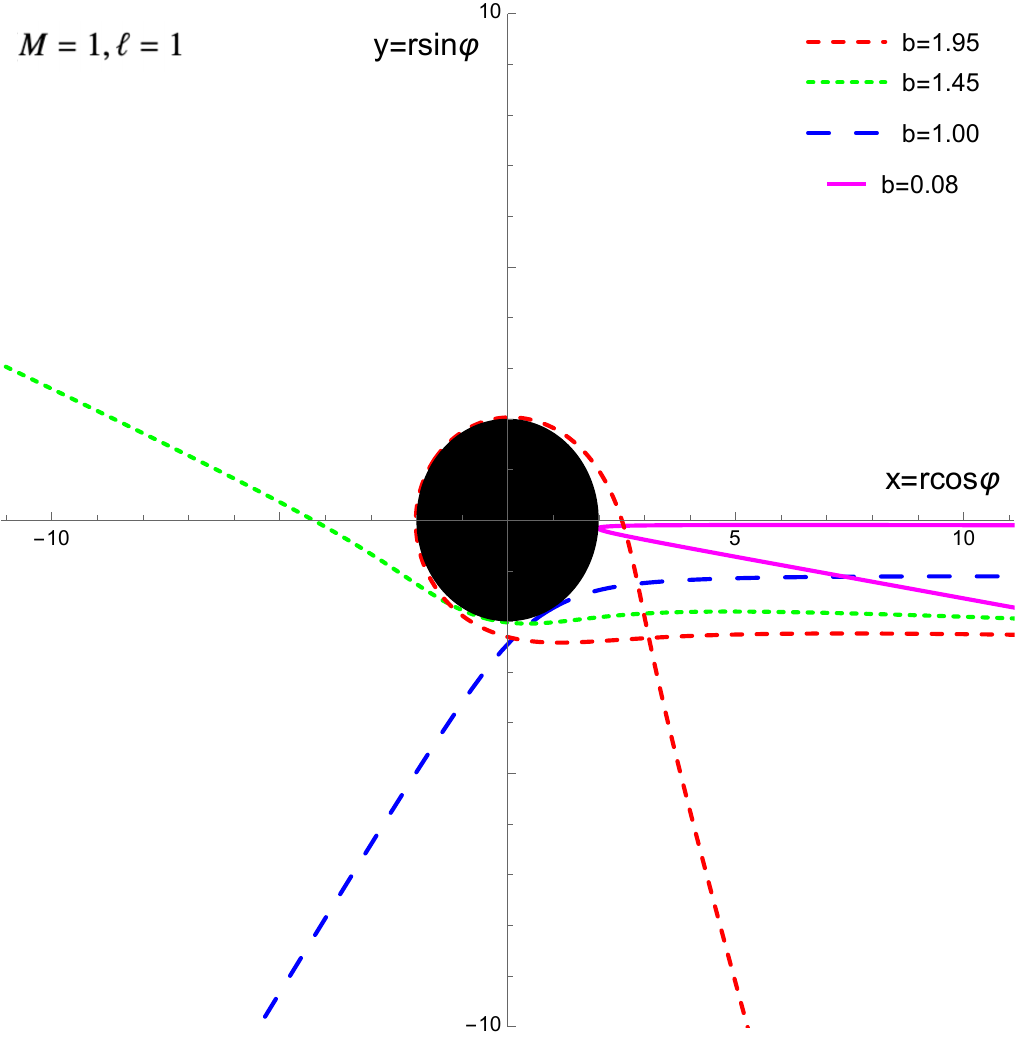}
\end{center}
\caption{For $b<2M$, the geodesics touch the CES and are reflected.}
\label{fig:2}
\end{figure}

Within the interval, $0<b<2M$, geodesics reach the CES with finite affine parameter and zero coordinate velocity, $\dot r=0$, and then get reflected at the CES. Integrating numerically the radial Carroll geodesic equation   
\eq{
\frac{\ddot r}{\ell^2} = \frac{M}{b^2r^2} + \frac{1}{r^3} - \frac{3M}{r^4}
}{eq:whatever}
with the initial data $r(0)=2M$, $\dot{r}(0)=0$ and the angular
equation \eqref{eq:cg14} with $\varphi(0)=0$ yields the orbits shown in
Fig.~\ref{fig:2}. We also created a \href{http://quark.itp.tuwien.ac.at/~grumil/Carroll_geodesics.nb}{Mathematica notebook} that displays geodesics as function of the impact parameter \cite{Mathematicanotebook}. 

In conclusion, Carroll black holes act like a perfect mirror for small impact parameters, unlike their Lorentzian counterparts, which in that regime absorb everything thrown at them. The reflection happens at the locus where the geodesics kiss the CES, providing congenial context to L\'evy-Leblond's allusion to Lewis Carroll's book ``Through the Looking Glass''.

\section{Conclusions}

Working intrinsically on a Carrollian geometry, we have constructed the Carrollian analog of the geodesic action \eqref{eq:cg3}. Remarkably, the equations of motion \eqref{eq:cg5}-\eqref{eq:cg8} lead to non-trivial dynamics: the Carrollian universe admits non-trivial orbits for test particles. 

\begin{table}
\begin{center}
\begin{tabular}{|l|l|}
\hline Value of $b$ & Key property of Carroll geodesics \\ \hline
\,$b\gg 2M$ & small deflection angle \eqref{eq:cg25} with $u_0=1/b$\\
\,$b>2M$ & finite deflection angle \eqref{eq:cg26} with $x=2M/b$ \\
\,$b\gtrsim2M$ & winding orbits \eqref{eq:cg41}, see purple/dashed orbit in Fig.~\ref{fig:1} \\
\,$b=2M$ & unstable circular orbit at CES or tangential orbit \eqref{eq:lalapetzNew}\,\\
\,$0\leq b<2M$\, & particle reflected by CES, see Eq.~\eqref{eq:whatever} and Fig.~\ref{fig:2} 
\\
\;$b^2<0$ & impossible unless $r<2M$ \\
\hline
\end{tabular}
\caption{Properties of Carroll geodesics for various values of $b$}
\label{tab:1}
\end{center}
\end{table}

Focusing on the Carroll--Schwarzschild background, we have seen that the radial geodesics are strikingly similar to the relativistic case, except that the energy enters the effective potential \eqref{eq:cg18}. This feature prevents circular orbits anywhere except at the CES, where such an orbit is unstable. We then computed the deflection angle, winding numbers, and showed that for small impact parameters $b$ \eqref{eq:angelinajolie}, the Carroll black hole acts as a perfect mirror. Table~\ref{tab:1} summarizes properties of Carroll geodesics with various ranges of the impact parameter $b$, displayed in Figs.~\ref{fig:1} and \ref{fig:2} (see also \cite{Mathematicanotebook}).

There are many outlooks of this paper, such as computing geodesics on different Carroll backgrounds, revisiting our results in the first order formulation, studying the phase space near the CES, applying our results to cosmology, exploring quantum effects like the Hawking effect and evaporation, etc. In conclusion, after many years and efforts in Carrollian physics, finding non-trivial motion could open new fascinating avenues to explore, and one cannot avoid thinking of the famous expression by Galileo, ``eppur si muove''.

\paragraph{Acknowledgements} 
L.C.~thanks Etera Livine for stimulating him to study simple Carroll questions. D.G.~thanks C\'eline Zwikel for hosting his research stay at  Perimeter Institute, where this project  started. We thank Ankit Aggarwal and Florian Ecker for comments on a nearly final version of this work. We are grateful to the organizers and participants of the kickoff event for the Simons Collaboration  on Celestial Holography, where some  discussions were finalized. 

Research at Perimeter Institute is supported in part by the Government of Canada through the Department of Innovation, Science and Economic Development Canada and by the Province of Ontario through the Ministry of Colleges and Universities. This work was supported by the Austrian Science Fund (FWF), projects P 32581, P 33789, and P 36619.

\bibliographystyle{uiuchept}
\bibliography{review}

\end{document}